%% file: 3C66A_ms.tex
\documentclass[12pt,preprint]{aastex}

\usepackage{lineno}
\usepackage{color}

\slugcomment{Accepted to ApJ February 8th, 2013}

\shorttitle{3C 66A Redshift and EBL}
\shortauthors{Furniss et al.}

\begin{document}
\linenumbers
\title{On the Redshift of the Very High Energy Blazar 3C 66A}

\author{A. Furniss\altaffilmark{1},
M. Fumagalli\altaffilmark{2,3,4},
C. Danforth\altaffilmark{5},
D. A. Williams\altaffilmark{1},
J. X. Prochaska\altaffilmark{6}
}
\altaffiltext{1}{Santa Cruz Institute of Particle Physics and Department of Physics, University of California Santa Cruz, 1156 High Street, Santa Cruz, CA 95064, USA.}
\altaffiltext{2}{Carnegie Observatories, 813 Santa Barbara Street, Pasadena, CA 91101, USA.}
\altaffiltext{3}{Department of Astrophysics, Princeton University, Princeton, NJ 08544-1001, USA.}
\altaffiltext{4}{Hubble Fellow}
\altaffiltext{5}{CASA, Department of Astrophysical and Planetary Sciences, University of Colorado, 389-UCB, Boulder, CO 80309, USA.}
\altaffiltext{6}{Department of Astronomy and Astrophysics, UCO/Lick Observatory, University of California, 1156 High Street, Santa Cruz, CA 95064, USA.}


\begin{abstract}
As a bright gamma-ray source, 3C\,66A is of great interest to the high-energy astrophysics community, having a potential for placing cosmological constraints on models for the extragalactic background light (EBL) and the processes which contribute to this photon field. No firm spectroscopic redshift measurement has been possible for this blazar due to a lack of intrinsic emission and absorption features in optical spectra.  We present new far-ultraviolet spectra from the Hubble Space Telescope/Cosmic Origins Spectrograph (HST/COS) of the BL Lac object 3C\,66A covering the wavelength range 1132 to 1800 \AA\ . The data show a smooth continuum with intergalactic medium absorption features which can be used to place a firm lower limit on the blazar redshift of $z\ge0.3347$. An upper limit is set by statistically treating the non-detection of additional absorbers beyond $z=0.3347$, indicating a redshift of less than 0.41 at 99\% confidence and ruling out $z\ge$0.444 at 99.9\% confidence.   We conclude by showing how the redshift limits derived from the COS spectra remove the potential for this gamma-ray emitting blazar to place an upper limit on the flux of the EBL using high energy data from a flare in October of 2009.
\end{abstract}


\section{Introduction}

Direct measurements of the extragalactic background light (EBL) are difficult due to strong foreground sources in our solar system (Zodiacal light) and the Galaxy \citep{hauser}.  If a direct measurement were possible, it would only reflect the current integrated state, leaving still the model-dependent task of extracting the time evolution.  These difficulties have been overcome through the use of extragalactic very-high-energy (VHE, E$\ge$ 100 GeV) gamma rays from blazars with known redshifts (e. g. \cite{aharonian2006,mazin2007,albert2008,gilmore2009,orr2011}), the most commonly detected type of VHE extragalactic source.  A blazar is a type of active galactic nucleus (AGN) that has a jet pointed towards the observer, and exhibits a highly polarized broadband spectrum from beamed, non-thermal emission processes.  


The energy-dependent absorption of gamma rays by the EBL softens the intrinsic VHE gamma-ray spectra emitted by extragalactic objects.  The details of the absorption depend on the shape of the EBL spectral energy distribution (SED) in the near-IR to optical band.  Additionally, the total power and the shape of the SED of the EBL is shown to vary strongly with redshift in the currently available models, such as \cite{dominguez,gilmore2012,kneiske,finke2010}.  To correctly account for the gamma-ray absorption, an accurate redshift of the VHE extragalactic target is required. 

Approximately one-third of the current VHE extragalactic catalog\footnote{\tt http://tevcat.uchicago.edu/} is made up by blazars at unknown or poorly-constrained redshift.  3C\,66A is one of these blazars, with an uncertain spectroscopic redshift based on possible corroborating measurements of single lines \citep{miller,kinney, bramel}.  Despite multiple attempts, in particular two high signal-to-noise measurements using Keck~I (shown in Figure 1), no solid spectroscopic measurement based on the detection of multiple lines from the host galaxy has been possible.  The lack of spectral features is not surprising given that 3C\,66A is a BL Lac-type active galactic nucleus that, by definition, displays weak or no lines. 

To overcome the inherent featureless characteristic of the 3C\,66A optical spectrum and enable deabsorption of the VHE spectrum with reliable redshift information, we have determined limits on the redshift of the blazar through the observation and statistical analysis of far UV (FUV) absorption by the low $z$ intergalactic medium (IGM). This method, already applied to the VHE blazars PG\,1553+113 and S5\,0716+714 \citep{danforth2010,danforth2013}, sets a redshift lower limit using absorption lines caused by the intervening IGM.  Further, given the expected distribution of IGM absorbers as a function of redshift, one can model any lack of absorption lines at longer wavelengths to statistically infer an upper limit on the blazar redshift.

\section{Observations and Spectral Analysis}
3C\,66A was observed with the Cosmic Origins Spectrograph (COS) during two epochs as part of two different programs.  The blazar was observed for five HST orbits on 1 November 2012, with the medium resolution G130M grating COS/G130M ($1135<\lambda<1450$ \AA\,, 15.3 ksec) as part of program 12621 (PI: Stocke). Three more orbits were devoted to observations with the COS/G160M ($1400<\lambda<1795$ \AA\,, 7.2 ksec) grating under program 12863 (PI: Furniss) on 8 November 2012.  The calibrated, one-dimensional spectra for each exposure were obtained from the Mikulski Archive for Space Telescopes (MAST).

The G130M data show a flux mis-match between the short and long-wavelength segment of each exposure and a $\sim8\%$ correction is applied to each before coaddition to bring them into the expected smooth continuum.  The G160M data are considerably noisier; no flux discrepancy was observed and no correction was undertaken.  The corrected exposures were then coadded with the standard IDL procedures described in detail by \cite{danforth2010}.  This procedure includes an automatic scaling of the exposures taken during different epochs.  The continuum flux level appears to have varied by $\la10\%$ during the week between observing epochs, well within the current flux calibration uncertainty.

The combined spectrum continuously covers the wavelength interval $1132-1800$ \AA\,, and shows the expected smooth continuum and narrow absorption features.  The data quality varies over the spectral range due to the different sensitivities and exposure times in the two gratings.  The mean signal to noise per pixel in the continuum is $\sim10$ ($\sim5$) with nominal dispersions of 9.97m\AA\,/pixel (12.23\AA\,/pixel) in the G130M (G160M) portion of the spectrum.   Signal to noise values per seven-pixel resolution element are approximately twice these values (see \cite{keeney2012}). For additional details on the COS instrument, see \cite{ghavamian} and \cite{kriss}.

Detailed analysis of these data and of the intervening absorption line systems will be presented in Danforth et~al. (in prep.). In this paper, we exclusively focus on the spectral features that are useful to constrain the unknown redshift of the blazar ($z_{\rm blazar}$).  The goal of the following analysis is to use absorption lines that arise from gas clouds in the IGM to establish a firm lower limit on the distance to 3C\,66A and to set an upper limit for the blazar redshift based on a statistical argument. 

A visual inspection of the spectrum reveals the presence of multiple absorption systems for which both Lyman$-\alpha$ and Lyman$-\beta$ (Ly$\alpha$ and Ly$\beta$) lines are detected. Among those, we identify three clouds at $z_{\rm abs}\sim 0.3283$, 0.3333, and 0.3347 (see Figure \ref{threesystems}). All other lines detected at $>$4 sigma significance redward of these three Ly$\alpha$ systems are identified as Milky Way absorption (see Figure \ref{redspectrum}).  Thus, because of the presence of a system at $z=0.3347$, we set a secure redshift lower limit of 3C\,66A at $z_{\rm blazar} \ge z_{\rm ll} = 0.3347$. We also search the spectrum for \ion{O}{6} ($\lambda\lambda 1031,1037$) 
doublets that, owing to their bluer rest-frame wavelengths, could yield a more stringent redshift lower limit than the one set by absorption in the Lyman series. However, we do not find any instances of absorption beyond $z\sim 0.33$.

Next, we can exploit the lack of absorption beyond $z_{\rm ll} = 0.3347$ to set an upper limit $z_{\rm ul}$ to the blazar redshift following a statistical argument. The frequency of absorption lines arising from the Lyman forest in the local universe has been measured along sightlines to extragalactic sources by different authors \citep[e.g.][]{penton2004,danforth2008}. It is common to express this quantity with the function $dN(W>W_{0})/dz$ which describes the average number of absorption lines with rest-frame equivalent width in excess to $W_{0}$ per unit redshift.  We can therefore estimate the number of lines we expect to detect between $z_{\rm ll}$ and $z_{\rm ul}$, given the rest-frame limiting equivalent width $W_{\rm lim}$ of the COS spectrum.  By comparing the predicted number of absorption lines in a given redshift interval with the lack of detection beyond $z_{\rm ll} = 0.3347$, we obtain redshift upper limit. 
 
First, we generate 1000 mock spectra in the observed wavelength range $1215-1800$ \AA\ by drawing Lyman forest lines from a distribution as a function of redshift such that the number of lines satisfies the observed $dN(W>W_{0})/dz$. In this analysis, we assume no evolution in Lyman-$\alpha$ forest line incidence and adopt the frequency distribution from \citet{danforth2008}, although a similar result is obtained if we adopted the distribution from \citet{penton2004}.  Next, we assign to each line a Doppler parameter 
drawn from the observed distribution in the local IGM \citep{danforth2008}.  During this step, we assume that the Doppler parameter is not correlated with the equivalent width of the line. Given a line equivalent width, its redshift, and a Doppler parameter, we compute the observed limiting equivalent width $W_{\rm lim}$ (at $5\sigma$) using the formalism developed for COS spectra by \citet{keeney2012} and we record only those lines which would be detected in the observed COS spectrum.  Note that this procedure naturally accounts for ``shadowing'' due to Milky Way absorption lines.

The top panel of Figure \ref{limitzmc} shows the number of intervening absorption lines detected in 1000 mock spectra within the redshift interval $0.335 \lesssim z \lesssim 0.444$. According to this Figure, we should expect to detect $\sim 5$ or more lines if 3C\,66A lies at $z_{\rm ul}>0.444$, and, although realizations with no lines are possible, they are extremely rare ($< 1 \%$ of the total trials). Under the simplistic assumption that the number of absorption lines is not correlated in velocity space, the mock realizations shown in the top panel of Figure 4 follow a Poisson distribution. Therefore, we can adopt Poisson statistics to express the probability of finding no detected lines between $z_{\rm ll}$ and $z_{\rm ul}$, given a typical number of Lyman forest lines in that redshift interval $N(z_{\rm ll} < z < z_{\rm ul})$.  

As shown in the bottom panel of Figure \ref{limitzmc}, the expected number of absorption lines increases proportionally to the redshift interval $\Delta z=z_{\rm ul} - z_{\rm ll}$ and the probability of finding no absorption lines $P(N=0)$ exponentially decreases with redshift.  At $z_{\rm ul} \sim 0.41$, $P(N=0)\sim0.01$, and therefore we conclude that 3C\,66A is likely to lie between $0.3347<z_{\rm blazar}\lesssim0.41$. We can further rule out $z_{\rm blazar}\gtrsim 0.444$ based on the 
fact that $P(N=0)\sim0.001$ for $z_{\rm ul} \sim 0.444$.  We note that consistent probabilities can be recovered directly from the Monte Carlo simulations, without explicitly using Poisson statistics.  However, it should be noted that our Monte Carlo simulations do not include correlated absorption systems in the Ly$\alpha$ forest.  Further, this calculation does not account for mechanisms that could enhance (e.g. galaxy clustering) or suppress (photoionization along the line of sight) the incidence of Ly$\alpha$ lines in proximity to a blazar compared to the mean value observed in the IGM, although there is no evidence for highly ionized gas ( i.e. \ion{N}{5} absorption) at $z\sim0.335$.   With proximity effects included, the predicted limits are subject to $\sim 1000~\rm km~s^{-1}$ uncertainty  (i.e. $\sim 0.003$ in redshift space).  Notably, there have been previous suggestions that 3C\,66A is a member of a cluster at $z \sim 0.37$ \citep{butcher,wurtz,clustering}.

The limits placing the redshift between 0.3347 and 0.41 disfavor the past tentative measurements of $z=0.444$ by \cite{miller} and \cite{kinney}, both of which were based on the measurement of single, weak lines.  The limits derived from the COS observations are, however, in good agreement with other past estimates of the blazar distance.  \cite{finke2008} set a lower limit of z $\ge$ 0.096, an estimation based on the expected equivalent widths of absorption features in the blazar host galaxy, while a distance estimate of $z\simeq0.321$, noting a large error, was formed based on the assumption that host galaxies of a blazars could be taken as standard candles.  An estimate for the blazar redshift of $z=0.34\pm0.05$ was found by \cite{prandini}, who extracted the approximate redshift by correcting the TeV spectrum of the blazar for EBL absorption to match the index measured by the \textit{Fermi} Large Area Telescope (LAT; \cite{atwood}), most sensitive to gamma rays between 300 MeV and $\sim$100 GeV which are largely unaffected by the EBL.  The redshift limits for 3C\,66A are also in good agreement with a recent EBL model-independent study of the gamma-ray horizon, as determined by synchrotron self-Compton modeling of VHE blazar broadband spectra \citep{dominguezCGRH}.

\section{Absorption of Very High Energy Gamma-rays from 3C 66A}
The energy- and redshift-dependent absorption of gamma rays by the EBL can be estimated using model-specific optical depths, $\tau(E,z)$, where the intrinsic flux ($F_{int}$) can be estimated by the observed flux ($F_{obs}$) using the relation $F_{int}\sim F_{obs}\times e^{\tau(E,z)}$.  The intrinsic index of a blazar can be used to estimate the spectral properties of the EBL under the physically motivated assumption that the intrinsic spectrum of a source undergoing Fermi shock acceleration, characterized by the power-law dN/dE $\propto E^{-\Gamma}$, cannot be harder than $\Gamma=1.5$.  If the intrinsic VHE spectrum is significantly harder than the $\Gamma$=1.5 limit, it can be argued that the gamma-ray opacity of the EBL model which was used for deabsorption is too high.  The index limit of 1.5 is derived from the standard leptonic and hadronic emission scenarios used to describe blazar non-thermal emission.  
This limit is also in agreement with the hardest gamma-ray index reported by the \textit{Fermi} LAT for a blazar \citep{2fgl}.  The indices for sources derived from photons with energies of less than 100 GeV are not significantly affected by EBL absorption and so reflect the intrinsically emitted spectra of blazars in the high energy gamma-ray band.  

Under the assumption that blazars do not harden with increasing energy, EBL flux constraints are also possible by comparing deabsorbed VHE spectra to the extrapolations based on the LAT-measured spectral indices.  
Using this method, the \textit{Fermi} and VERITAS indices measured during a state of elevated flux from 3C\,66A in October of 2009 ($\Gamma=1.8\pm0.1_{stat}$ and $4.1\pm0.6_{stat}$, respectively; \cite{abdo3C66A}) allow the investigation of possible constraints on the EBL density, pending a reliable distance measurement.  Previously, the deabsorption of the VHE spectrum of 3C\,66A has been completed with the uncertain spectroscopic redshift of $z=0.444$ (e. g. \cite{finke2010,dominguez,aleksic}).  Notably, \cite{gilmore2012} shows that the intrinsic spectrum derived from deabsorption of 3C\,66A with the tentative redshift of $z=0.444$ is the hardest of the deabsorbed VHE BL Lacertae objects.

Figure 5 shows the VERITAS measured VHE spectrum of the blazar 3C\,66A from \cite{abdo3C66A} (black solid line) when deabsorbed for the redshift upper and lower limits from this work.  These deabsorbed spectra are calculated by multiplying the measured differential flux values by $e^{\tau(E,z)}$ for various EBL models.  The resulting intrinsic flux estimates are then refit with a differential power-law for the redshift lower limit (top) and 99\% upper limit (bottom).  The fitted intrinsic indices for both the lower and upper limits on redshift are summarized in Table 1. The hardest deabsorbed spectra result from the \cite{finke2010} EBL model, but all fitted power-laws provide indices softer than the $\Gamma=1.5$ limit (shown for reference in Figure 5 by the grey solid line at a comparable normalization to the deabsorbed spectra).  The resulting indices are also below the \textit{Fermi}-LAT measured index of $\Gamma$=1.8$\pm$0.1.

\section{Conclusion}
Observation of the $z\sim0$ Ly$\alpha$ forest in the direction of the 3C\,66A with HST COS provides a direct lower and statistical upper redshift limit for the blazar.  The detection of three clouds at $z_{\rm abs}\sim 0.3283$, 0.3333, and 0.3347 provide the $z=0.3347$ lower limit on the blazar redshift.  Assuming that the incidence of Lyman absorption systems is a Poisson distribution in $z$, we can conclude that the blazar is likely to lie $z_{\rm blazar}\lesssim0.41$ (99\% confidence level) and exclude a $z\ge0.444$ at 99.9\%. 

Based on the assumption that the intrinsic index cannot be harder than $\Gamma=1.5$, the redshift limits derived from the FUV observations do not place the blazar at a sufficient distance to utilize the observed VHE spectrum during an elevated state in October of 2009 to constrain the EBL density.  Moreover, the distance is not sufficient to extract an upper limit on the EBL density based on the similar assumption that the intrinsic VHE index is not harder than the \textit{Fermi} observed index.


\acknowledgments
We are grateful to Robert da Silva for insightful discussions on the statistical analysis of these data. Support for program HST-GO-12863 and for Hubble Fellow M.F. (grant HF-51305.01-A ) were provided by NASA awarded through grants provided by the Space Telescope Science Institute, which is operated by the Association of Universities for Research in Astronomy, 
Inc., for NASA, under contract NAS 5-26555.  Additional support for this work came from National Science Foundation award PHY-0970134.  C. D. was supported by NASA grants NNX08AC146 and NAS5-98043 to the University of Colorado at Boulder.

{\it Facilities:} \facility{HST (COS)}.

\clearpage

\begin{figure}
\includegraphics[scale=0.6,angle=90]{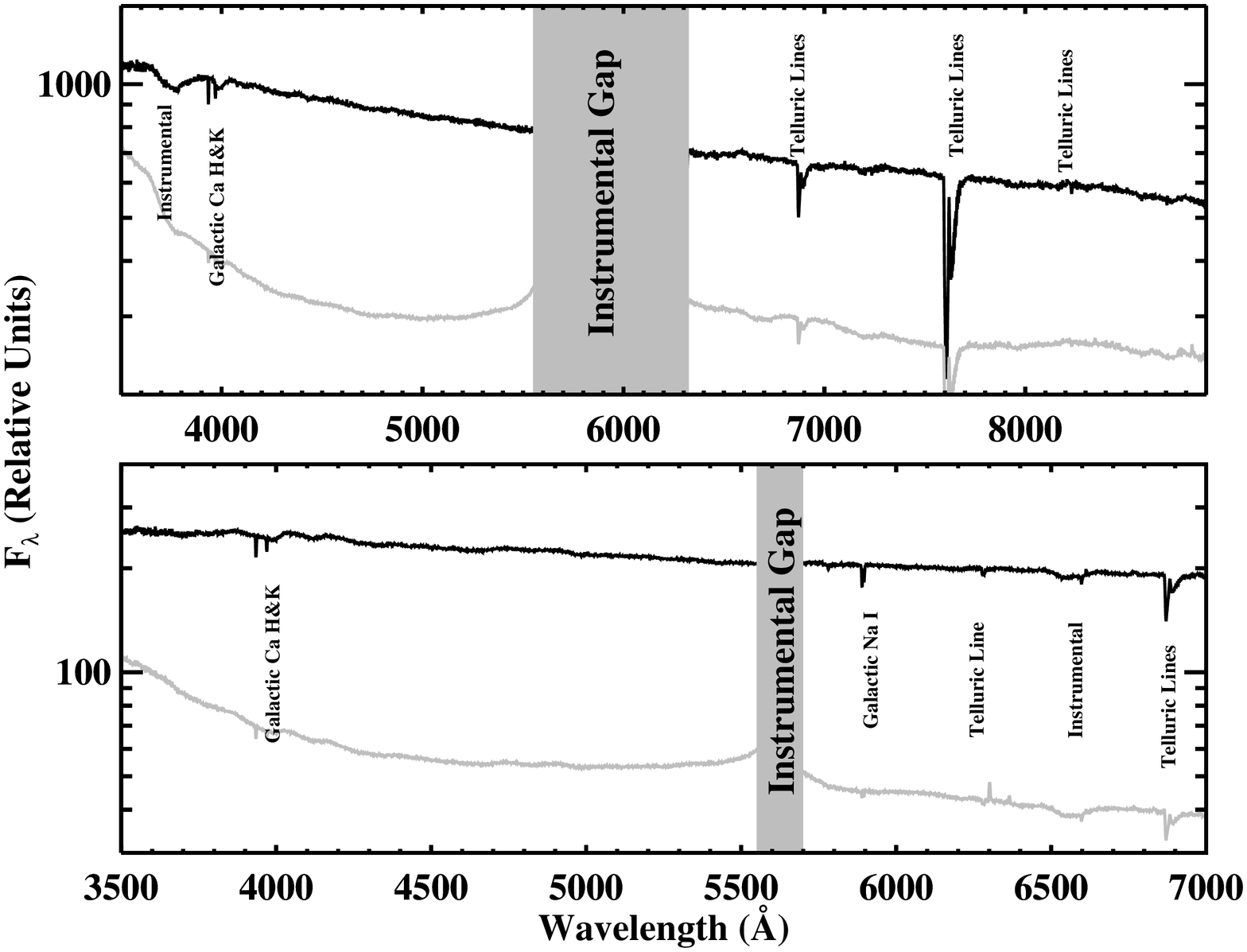}
\caption{Keck/LRIS spectra (black) and error array (scaled by 50$\times$; grey)
of the optical emission from 3C\,66A from September 2009 (top, relative high state) and October 2011 (bottom, relative low state).  The 
gaps in the spectra are due to the dichroic filter of the instrument.  We have additionally cut the 2011 spectrum at 7000 \AA\, due to uncertainties introduced in calibration.  All significant
absorption features identified in the spectra are associated with the Earth
or Milky Way.  The details of this spectral analysis for each of these observations are completed as described in  \cite{abdo3C66A}.  Even at this exquisite S/N (over 100 per pixel for both exposures) there are no 
features with which to place a constraint on the redshift of this blazar.
\label{fig_3c66a} }
\end{figure}

\begin{figure}
  \centering
  \includegraphics[scale=0.5,angle=90]{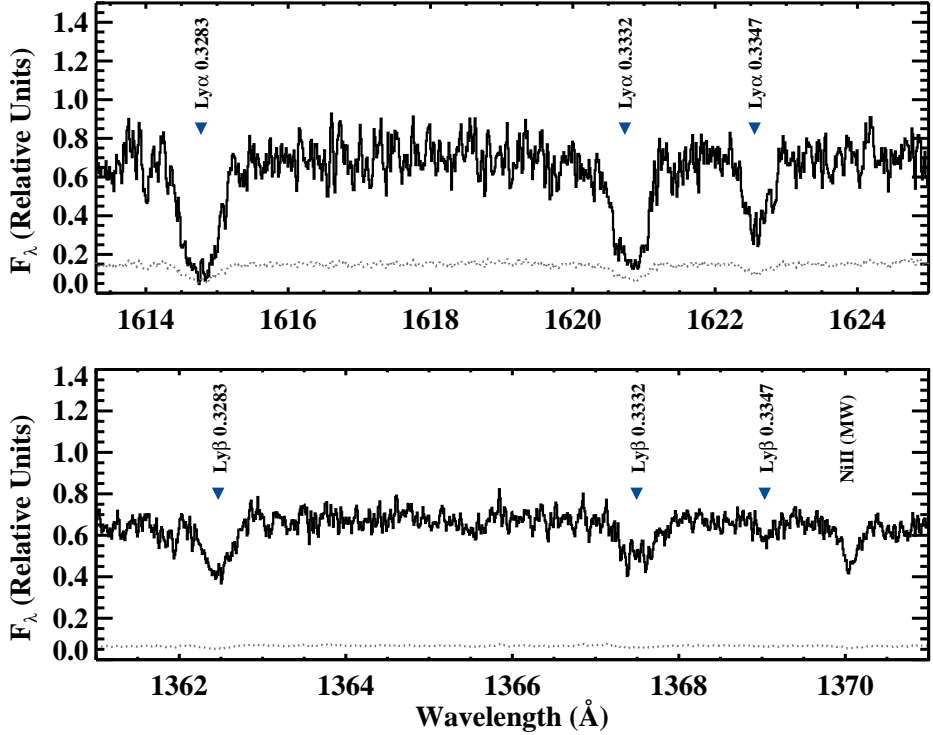}
  \caption{Detail of the COS spectrum of 3C\,66A in the regions where we identify 
  Ly$\alpha$ (top) and corresponding Ly$\beta$ (bottom) absorption lines for three gas clouds 
  at $z_{\rm abs}\sim 0.3283$, 0.3333, and 0.3347. Absorption associated with 
  Galactic \ion{Ni}{2} is also labeled in the bottom panel.}\label{threesystems}
\end{figure}

\begin{figure}
  \centering
  \includegraphics[scale=0.5,angle=90]{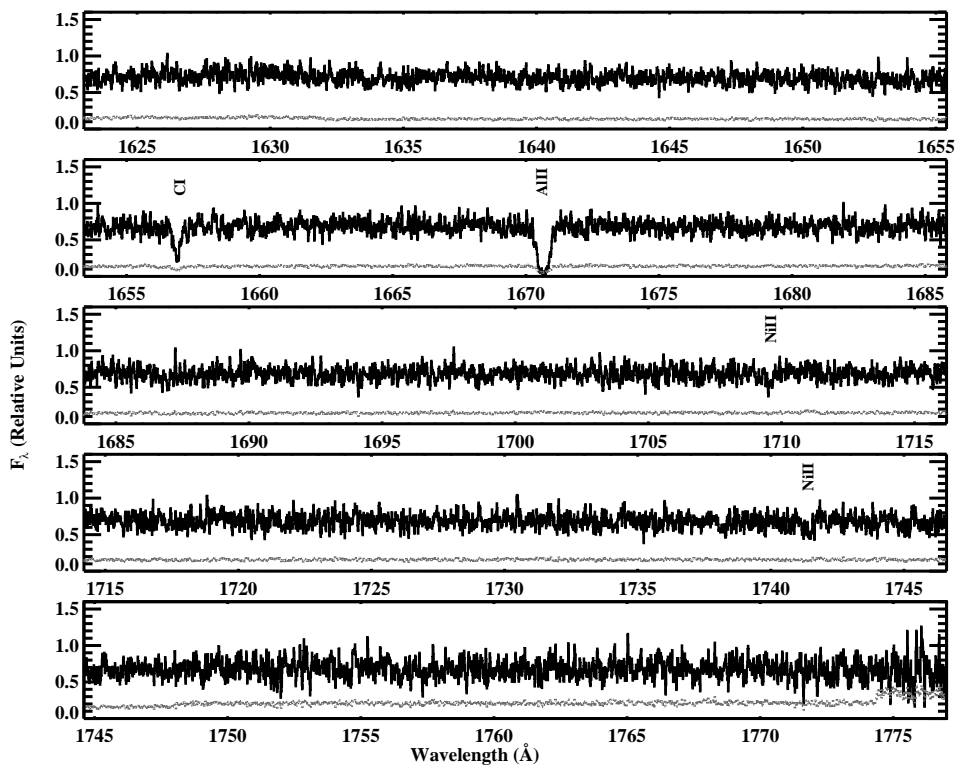}
  \caption{Red portion of the G160M spectrum, redward of where we identify Ly$\alpha$ lines
    at $z_{\rm abs}\sim 0.33$. All the labeled lines arise in the Milky Way. The lack
    of absorption of extragalactic origin places the redshift lower limit of 3C\,66A at 
    $z_{\rm blazar} \ge 0.3347$.}\label{redspectrum}
\end{figure}


\begin{figure}
  \begin{tabular}{c}
    \includegraphics[scale=0.32,angle=90]{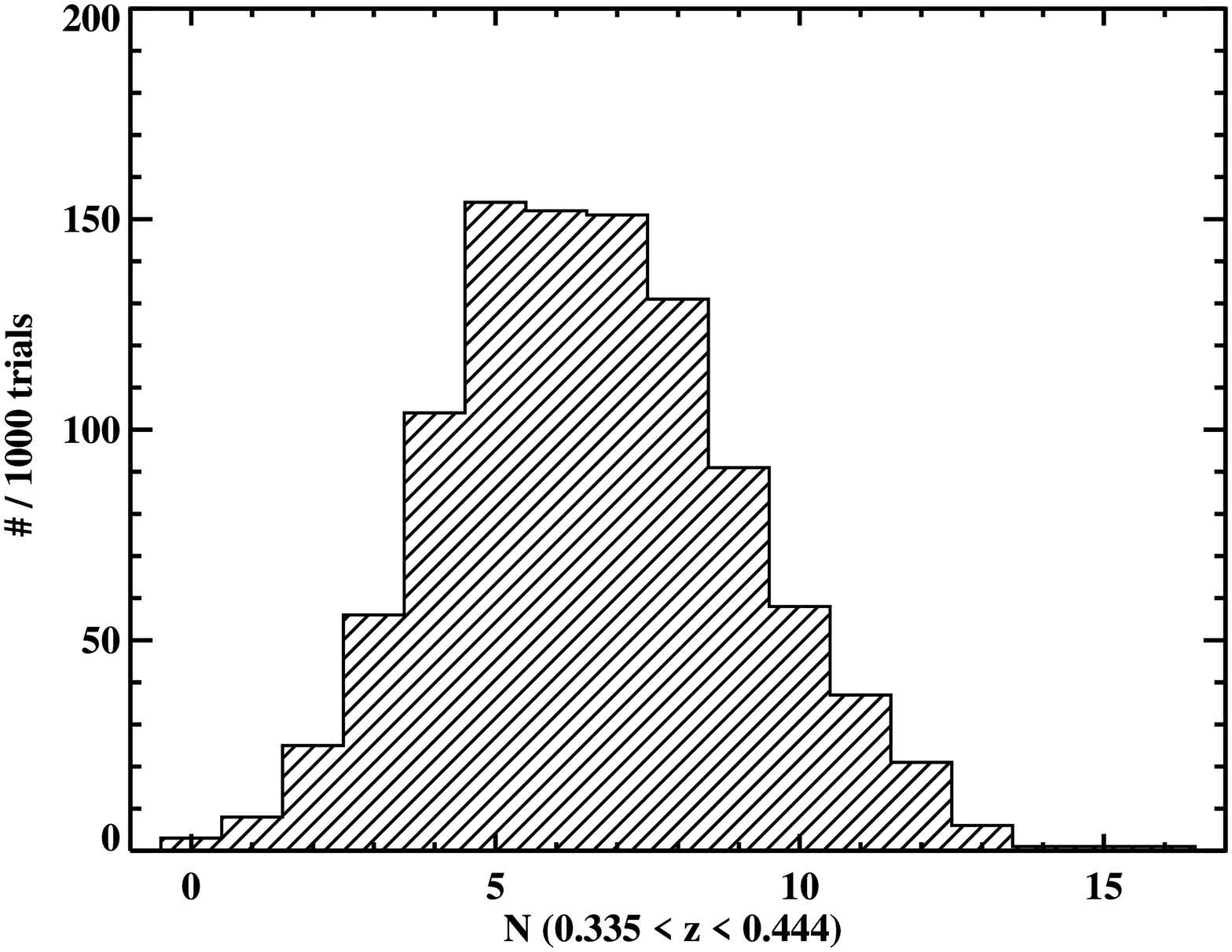}\\
    \includegraphics[scale=0.32,angle=90]{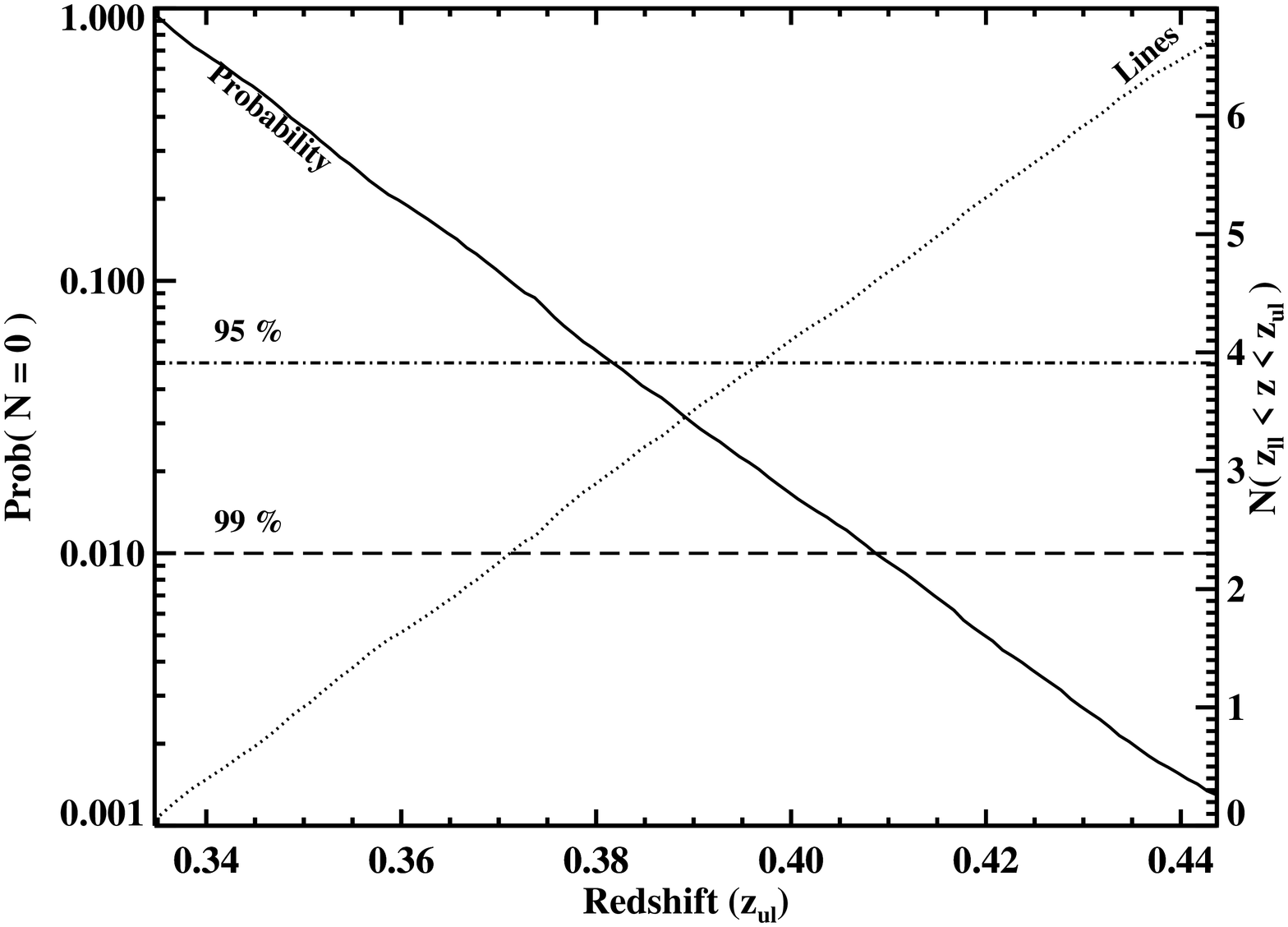}
  \end{tabular}
  \caption{{\it Top} Distribution of the number of lines detected in 1000 mock spectra for 
    $0.335 < z < 0.444$.  {\it Bottom} The probability to observe 
    no Ly$\alpha$ lines if 3C\,66A lies beyond $z_{\rm ul}$ given the expected number of Ly$\alpha$ 
    lines in the redshift interval $z_{\rm ll} < z < z_{\rm ul}$ derived from Monte Carlo simulations
    (dotted line).}\label{limitzmc}
\end{figure}

\begin{figure}
  \centering
  \includegraphics[scale=0.8]{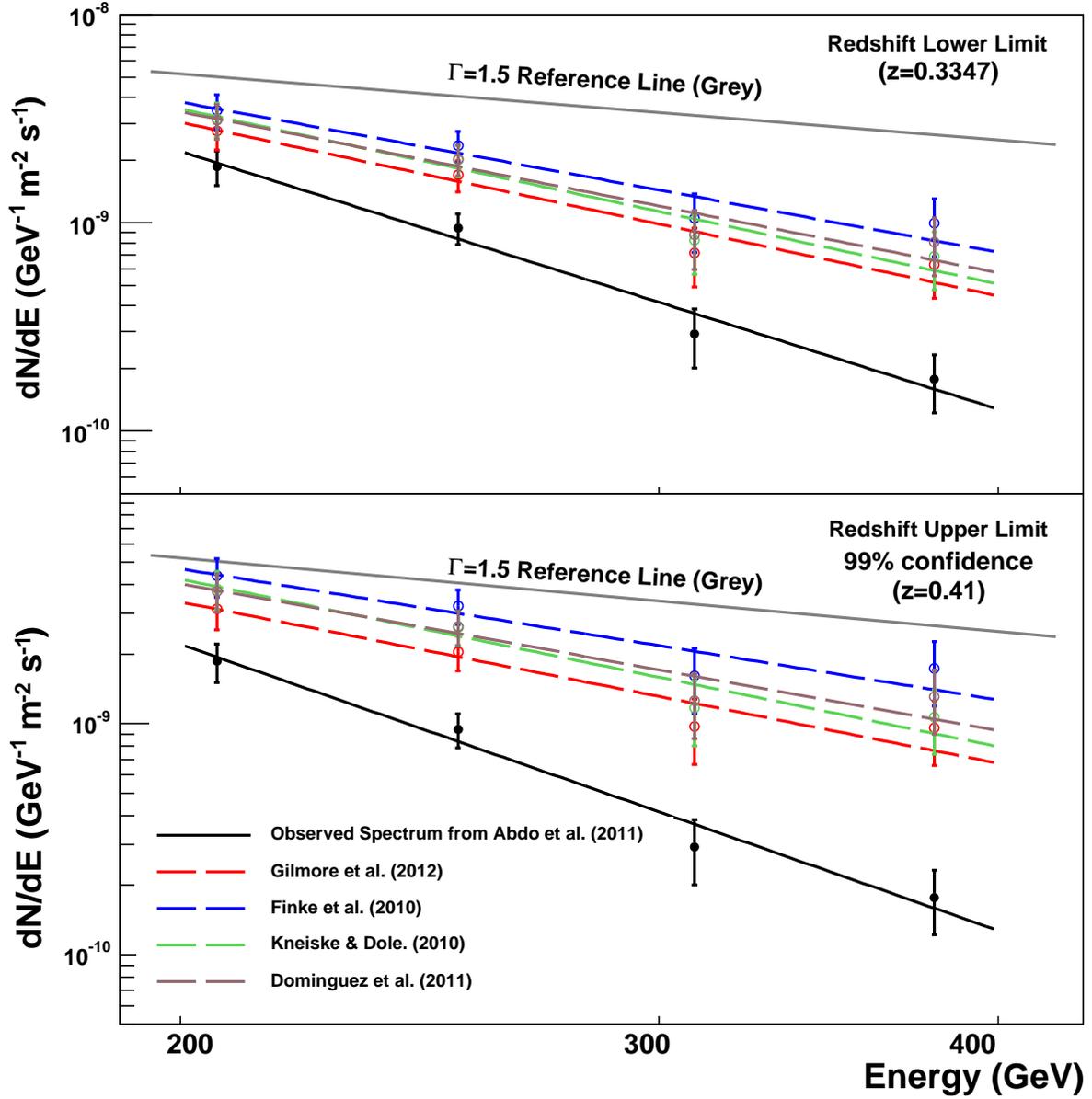}
  \caption{Deabsorbed spectra for 3C\,66A for the $z_{\rm ll}$ of 0.3347 (top panel) and 99\% confidence level $z_{\rm ul}$ of 0.41 (bottom panel), where the observed VHE spectrum (black solid line) is taken from \cite{abdo3C66A}, with an index of $\Gamma=4.1\pm 0.6_{stat}$ for the applied differential power-law of the form $dN/dE=(E/E_o)^{-\Gamma}$.  For reference, a spectrum with an index of $\Gamma$=1.5 is shown as the theoretical limit for an intrinsic index, as explained in the text.  The resulting indices for each redshift and model are summarized in Table 1. }\label{deabsorbed}
\end{figure}

\input{indices.tex}

\end{document}

%% file: indices.tex
\begin{deluxetable}{lcc}
\tabletypesize{\scriptsize}
\tablecaption{Intrinsic indices ($\Gamma$) resulting from the deabsorption of the VERITAS observed spectrum reported in \cite{abdo3C66A}.  Indices are calculated by taking the VERITAS-measured differential flux and flux errors and multiplying by $e^{\tau}$, where $\tau$ is an energy and redshift dependent optical depth taken from the EBL models.  The resulting flux in each bin is then fit with the differential power-law of the form $dN/dE=(E/E_o)^{-\Gamma}$, where $E_o$ is 250 GeV.}
\tablewidth{0pt}
\tablehead{
  \colhead{EBL}&  \colhead{Deabsorbed}&  \colhead{Deabsorbed} \\
  \colhead{Model}&  \colhead{Index}&  \colhead{Index} \\  
  \colhead{Used}&  \colhead{$z=0.3347$}&  \colhead{$z=0.41$} \\}    
 \startdata
\cite{gilmore2012}&2.8$\pm$0.6&2.3$\pm$0.6\\
\cite{finke2010}&2.4$\pm$0.6&1.9$\pm$0.6\\
\cite{kneiske}&2.8$\pm$0.6&2.4$\pm$0.6\\
\cite{dominguez}&2.6$\pm$0.6&2.1$\pm$0.6\\
 \enddata
\end{deluxetable}